\documentclass[12pt]{article}

\begin{document}
\input{epsf.tex}
\title{On the possible critical behaviour
       of a marginally stable stellar disc}

\author{E.~I.~Ivannikova and  M.~N.~Maksumov\\
{\em Institute of Astrophysics,  Tajik  Academy of Sciences,}\\
{\em Dushanbe, Republic Tajikistan}}


\maketitle

\begin{abstract}
Using hydrodynamic approach, it is shown that the properties of a
marginally stable
collisionless
stellar disc resemble those of a thermodynamic
system undergoing a
gas--liquid
phase transition.  The  maximum in
Toomre's stability  diagram, which separates gravitationally stable
and unstable states with respect to  axisymmetric perturbations,  can
be treated   as  a  critical point
for
this transition. Static
perturbations of stellar density are explored  and the  mean
perturbation amplitude is considered as the order parameter of the
theory.
The disc's state is assumed to change
 as the disc passes through
the critical point.  Since the disc  tends to retain hydrostatic
equilibrium, structures can be formed spontaneously,
identifiable with a seed spiral structure.
A power-law scaling of the order  parameter  in  the
vicinity of the critical point has been found. The susceptibility and
other Landau--Weiss exponents similar to those in the Van  der  Waals
theory are
calculated. The critical behaviour of marginally stable discs at the
initial stage of their evolution occurs in numerical simulations
where snapshots of stellar positions reveal stellar splinters and
crescents diverging from the disc centre. These structures can be a
result of the phase transition.
In numerical simulations, these structures eventually reduce to
decaying worm-type features because of the `heating'
most likely resulting from instability of stellar orbits due to
resonances. Under favourable conditions  the critical behaviour
leading to the
establishment of order in a stellar disc can result in the generation
of a spiral structure.

\end{abstract}


\section{Introduction}
Our aim is to clarify whether an ordered state can arise
spontaneously in a marginally stable stellar disc. We suggest that
the marginal state can be a critical one with respect to
a phase transition. We show that a
self-gravitating
galactic stellar disc is most  sensitive  to static
perturbations of gravitational field at a  certain length scale when
the disc is in a marginal state (where chaotic stellar velocity is
equal to the local gravitational
momentum $\pi G \sigma/\kappa$, i.e., the mean gravity force times
the epicycle period, where $G$ is Newton's gravitational constant,
$\sigma$ the surface mass density and $\kappa$ the epicyclic
frequency).
Perturbations of stellar density
grow with the difference
between the radial velocity
dispersion and its Toomre's characteristic value, and
can modify the symmetry of the disc
when their scale is comparable to the size of the stellar epicycle,
the natural wavelength of oscillations.

The susceptibility of  the stellar `gas' is singular at the top
point of the well-known Toomre's stability diagram (Toomre 1964).
Moreover, the static response  to small  perturbations of
the gravitational potential can be singular
in the region of
instability in the diagram. Therefore  internal physical
equilibrium of a self-gravitating disc requires
an analysis deeper than that
in terms of the standard  theory  of  dynamic
instability.

Guided by physical considerations (Saslaw 1987), we
assume that the marginal state of the disc is similar
to a critical
point of the gas--liquid phase transition  (as  a  special  case  of the
phase transition of the second kind), and the disc's  state
in unstable region
is similar to the condensed state of
the van  der  Waals  gas.
(We stress that this is just a formal similarity; the physics of the
stellar disc is quite different from that of a gas-liquid system.) Note,
however, that we consider not the loss of
thermodynamic stability as density increases, but rather the
occurrence of a susceptible  state, compressible and `elastic'
simultaneously.  Therefore, the physical properties of
perturbations in  the `van der Waals  region' are analyzed using
a nonlinear dependence of `pressure' perturbations on stellar density
perturbations.  We demonstrate the
possibility of a quasi-static compression of the perturbed stellar
`gas', obtain an equation for the order parameter (viz.\ the amplitude
of the static perturbations in  the stellar spatial density), and
investigate the perturbations in the neighbourhood  of the critical
point in Toomre's diagram.

In fact, a stellar disc is a collisionless dynamic system. The `elasticity'
of such a self-gravitating system results from  both the epicyclic
motions of stars under small
perturbations and the centrifugal `elasticity' of
the reference (circular) stellar orbits
themselves.

Although critical fluctuations can be important because of an
anomalous behaviour of  the susceptibility (Ivannikova, Ivanov \&
Maksumov 1994, 1997), we shall restrict ourselves to Landau's
approach and consider the possibility, for a self-gravitating stellar
system, to reach a new local equilibrium by developing an
ordered state. This approach uses an order parameter as a measure of
evolution in the disc structure as it undergoes a phase transition,
but fluctuations are ignored. Below we use the mean amplitude  of
density perturbations in the marginal disc as the order parameter. We
do not use the method of thermodynamic potentials because of the
well-known complications in its application to self-gravitating
systems (Saslaw 1987).

As follows Saslaw's arguments (\S\S 29.1,  29.3,  30.3, and
31.1 in Saslaw 1987), a direct introduction of gravity into the theory of
the van der Waals gas leads to a poor approximation, especially
because of the collective nature of the gravitational interaction.

The method used here combines the mean field approach, similar
to the mean molecular field theory of  Weiss (\S9.2 in Balescu 1975),
and the ideas related to  hydrostatic equilibrium  in an
external field and Landau's theory of the gas--liquid phase transition.
 This
differs from the fluctuational approach as the perturbations
 are considered to be ordered in space. It is
 assumed that
all the stars are identical, and so each equally interacts with the other
stars in the
disc. This interaction is described by the mean  field. This method
is not free
of  disadvantages similar to those of the theories of van der Waals,
 Weiss and Landau.

In Section 2, density perturbation in the pre-critical state is obtained
using linear  collisionless hydrodynamics (theory of
density waves, Lin \& Shu 1964) and then its limiting value
corresponding to the top point of Toomre's diagram is taken.
This approach is equivalent to the local gravithermodynamics,
as shown by Saslaw (1987, \S 31) for the Jeans criterion as an
example.

But the direct analysis of intrinsic (or proper internal)
and forced static perturbations
(as opposed to fluctuational ones)
at the critical point in terms of general thermodynamic relations
in Section 3
apparently provides the most straightforward explanation of the new
phase in the marginally stable disc.

Our results, albeit schematic, clarify an important  question as to how a
self-gravitating disc will respond to a static spiral
perturbation   of gravitational potential (in the approximation of tight
winding)
whose wave number is equal to the inverse Coriolis radius and how
the amplitude of
stellar density perturbations will vary if the disc is close to the marginal
stability (i.e., where the mean chaotic stellar velocity is comparable to the
mean gravitational momentum acting on stars that move along
the Coriolis circles). We assume that the response
of the disc is similar to a phase transition  of the second  kind,
and we use this assumption to suggest the form of the `pressure'
perturbations
in the post-critical regime.
 The marginal  state
appears to be critical with respect to compression driven by the spiral
perturbation of the gravitational potential, which can be expressed as
a growth  of
the susceptibility of the stellar density perturbations. We recall
that  the  critical point of the gas--liquid phase transition is
characterized by the
loss of thermodynamic stability as density increases.

    The stellar disc in our theory  does not have the  van
 der  Waals  equation  of  state. We found it difficult to obtain any
explicit form of the equation of state for the stellar disc. However,
it is sufficient for our purposes to know the `pressure'
perturbation alone. Our main tools are Toomre's stability diagram,
the  density
 wave  theory, and the theory of critical phenomena due to van  der  Waals,
  Weiss
 and  Landau. Important for the method used are Toomre's critical
 dispersion
 (as an indicator of criticality), the mean field approach borrowed
 from the theory of stellar collective phenomena,  and the ideas related
 to hydrostatic equilibrium in an external field.

We further assume that, in the critical region, the disc evolves into a
new equilibrium state similarly to the van der Waals gas.  Therefore we
consider here static perturbations in the stellar disc. The amplitude
of the perturbations is determined from equilibrium equations
in a constant external and internal gravitational fields. We recall
that Landau considered the thermodynamic potential for a fixed value
of the order parameter determined from the minimum condition for the
potential.

The concept of the critical point is modified because of the
long-range nature of the gravitational force and the absence of
stellar collisions (Saslaw 1987). An important role here is played by
Toomre's velocity dispersion, that is by the collective local
interaction, rather than by pairwise interparticle interactions of
van der Waals. The `centrifugal elasticity' also results in a
nonlinear term in the `pressure' perturbation.
We consider spontaneous condensations (`convergence')
 and rarification (`divergence') of stellar orbits in radius (similar
 to compressible perturbations in the van  der  Waals  gas),
but do not include such effects as `heating'.

\section{The response function, coherence length and
    susceptibility in the pre-critical region }
In this section we evaluate the response function using linear collisionless
hydrodynamics including both forced and internal (proper) perturbations in the
stellar spatial density.

Consider a nonaxisymmetric  perturbation of the stellar surface
density $\sigma_m$ (the order parameter), whose background value
is $\sigma$, induced by small perturbations of gravitational
potential $\psi_1=\psi_{1,\rm  int}+\psi_{1,\rm ext}$ (positive
definite), where the potential is represented as a sum of
internal, self-consistent field and external field. In the limit
of tight winding, we have (see Appendix A)
\begin{equation}
 \label{eq1}
\frac{\sigma_m}{\sigma}
 =-\frac{c^2_r\sigma_m/\sigma-\psi_1}{\kappa^2-{(\omega- m\Omega)}^2}\, k^{*2}\;,
\end{equation}
where $c_r$ is the radial velocity dispersion, $\Omega$ is the angular velocity of the
 disc, $\kappa$ is the epicyclic frequency, $\omega$ is the angular frequency of the
 perturbations, and $m/r$ and $k^*$ are the azimuthal and total wave numbers,
 respectively. Using $\psi_{1,\rm int}=2\pi G \sigma_m/|k^*|$, Eq.~(\ref{eq1})
 can be rewritten as
\begin{equation}\label{eq2}
    \frac{\sigma_m}{\sigma} [\kappa^2-(\omega-m\Omega)^2]
    = (- c^2_r k^{*2} + 2\pi G\sigma|k^*|)
        \frac{\sigma_m }{\sigma} + \psi_{1,\rm ext}k^{*2}.
\end{equation}
Assuming that $\omega-m\Omega=0$ (cf.\ Saslaw 1987, \S 31)
and dividing by $\kappa^2$, we obtain
\begin{equation}\label{eq2p}
\left( 1  +  { c^2_r k^ { * 2 } \over { \kappa^2 }}  -  {{ 2\pi G
    \sigma|k^*|} \over { \kappa^2 }} \right) \frac {
     \sigma_m } { \sigma } =  {{ \psi_ { 1, \rm ext }  k^ { * 2 }} \over {
     \kappa^2 }}\; .
\end{equation}

Consider the density response (positive definite) and complete the
square in $k^*$ on the left-hand side:
\[
 1+\frac{c^2_r k^{*2}}{\kappa^2}-\frac{2\pi G\sigma}{c_r\kappa}\,
 \frac{c_r|k^*|}{\kappa} +\frac{\pi^2 G^2\sigma^2}{c^2_r\kappa^2}
-\frac{\pi^2 G^2\sigma^2}{c^2_r\kappa^2}
\]
\begin{eqnarray*}
\mbox{}\qquad\qquad
&=&1+\left(\frac{c_r k^*}{\kappa}-\frac{\pi G\sigma}{c_r\kappa}\right)^2-
     \frac{c_{\rm T}^2}{c_r^2}\\
&=&1+\frac{c^2_r}{\kappa^2}
    \left(k^{*}-\frac{\pi G\sigma}{c_r^2}\right)^2
-\frac{c_{\rm T}^2}{c_r^2}\\
&=&\frac{c^2_r-c_{\rm T}^2}{c^2_r}+\frac{c^2_r}{\kappa^2}\,k^2\;,
\end{eqnarray*}
where we have taken into account that $k^*-\pi G\sigma/c_r^2=k$.

We can introduce a characteristic length (analogous to the correlation length)
quantifying the spatial coherence
of the perturbations in  the mean self-consistent
field. Two quantities with the dimension of length can be identified
 in the above expression: $r^2_{\rm corr}=c^2_r/\kappa^2$ and
 $L^2_{\rm corr}=r_{\rm corr}^2 c^2_r/(c^2_r-c^2_{\rm T})$.
 In terms of $L^2_{\rm corr}$, Eq.~(\ref{eq2p}) reduces to
\begin{equation}\label{eq2pp}
    \frac { \sigma_m } { \sigma } = {{ \psi_ { 1, \rm ext } \over { c^2_r }}
     {{ k^ { * 2 }} \over { L^ { -2 } _ {\rm corr }  +  k^2 }}} .
\end{equation}

This analysis resembles  the  theory  of Lin and Shu of spiral
 density waves  in its hydrodynamic form. But there is a crucial
 distinction.  We do not address the question of instability
 controlled by the magnitude of the radial velocity dispersion.
 Instead, we consider
two types of response in the stellar density and its scaling
with the velocity dispersion: a spontaneous response not affecting the disc
 energy and a forced response.

Having accepted $L_{\rm corr}$ as the coherence length, we see that the critical
exponent $\nu$ is equal to $1/2$, as it should in Landau's
theory;
from the definition of $\nu$,
$L_{\rm corr}\sim(c_r^2-c_{\rm T}^2)^{-\nu}$.
For  $c_r >c_{\rm T}$, i.e., in the pre-critical range of $c_r$, $L_ {\rm
 corr}$ is close to the stellar Coriolis radius as obtained by Ivannikova et al.\
 (1997) from different arguments. However, the expression obtained here can be
 extrapolated to the critical region.

\epsfxsize=4.0cm

\begin{figure}
\centerline{\epsfbox[50 0 150 100]{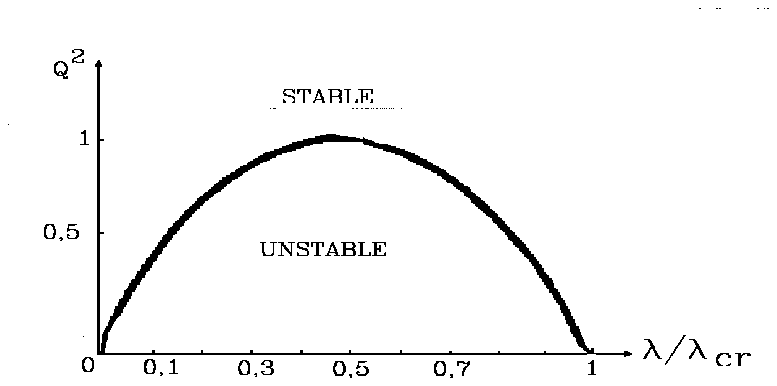}}
 \caption{Toomre's
stability diagram  for a thin stellar disc in the
 hydrodynamic approximation. The region of instability lies below the
marginal curve.  $Q$ is Toomre's parameter, $\lambda=2\pi/k$, and
$\lambda_{\rm cr}\equiv\lambda_{\rm T}.$}
\end{figure}


$L_{\rm corr}$ can also be identified with the spatial extent of the
pattern since its inverse is the width of the spread in the wave
 number space, and the expression in  parentheses on the left-hand
  side of Eq.~(\ref{eq2p}), if equated to zero, yields Toomre's
 dispersion relation for Lin's spiral density waves in the
 approximation of collisionless hydrodynamics with
 $\omega-m\Omega=0$. This is the equation of the marginal curve shown
 in Fig.~1.

Now consider the response in the vicinity of the top point in
 Toomre's diagram.  Assuming $k=0$ in Eq.~(\ref{eq2pp}), i.e.,
 $k^*=k_{\rm cr}=\pi G\sigma/c_r^2$, we obtain for the amplitude of a
 mode whose scale is close to the critical value $k_{\rm cr}^{-1}$:
\begin{equation}\label{eq2ppp}
 \frac{\sigma_m}{\sigma}
 = {{ \psi_ { 1, \rm ext } \over { c^2_r }} \, {{ c_{\rm T}^2 } \over {
 c^2_r-c_{\rm T}^2 }}} \;.
\end{equation}

This yields the  critical exponent as $\gamma=1$ for $c_r >c_{\rm T}$
 from the definition $\sigma_m=\chi\psi_{ 1,\rm ext}$ of the
 susceptibility $\chi$ and the definition of $\gamma$,
$\chi\sim(c_r^2-c_{\rm T}^2)^{-\gamma}$. As expected, the correlation
length is related to
the susceptibility (Ma 1976, Patashinsky \&
Pokrovsky 1975). The spontaneous value of the order parameter
$\sigma_m/\sigma$ in the
post-critical
state is also related to
$L_{\rm corr}$  (Patashinsky \& Pokrovsky 1975), i.e., is determined
not only by the direct local correlation, but also by the global
correlation chain whose scale grows as the critical point $c_r=c_{\rm
T}$ is approached.

 There are further implications of Eq.~(\ref{eq2pp}).  For
 $c_r<c_{\rm T}$, this equation is inapplicable if $k$ tends to  zero
 remaining on the marginal curve because then Fredholm's alternative
 for the inhomogeneous problem is violated. Physically this means
 that matter motions in the disc must be allowed, for which nonlinear
 analysis would be required.

Two types of the critical behaviour of the disc can occur.
Firstly, the disc can become dynamically unstable, leading
to processes similar to those studied by theory of synergetics.
Secondly, a phase transition can occur, similar to
the gas--liquid  phase  transition  at~ a  critical point.
Here we investigate the latter posibility.

\section{The order parameter, susceptibility and critical exponents
in the critical and post-critical states }
In the previous section  and Appendix, the disc  was  treated  in
the linearized WKB approximation as an isothermal disc. In this section
 a different, non-linear expression for the `pressure' perturbation
  must be used.
The reason  is that the properties of the disc matter change at
 $c_r=c_{\rm T}$.
The disc has different properties in the pre-critical state
 ($c_r>c_{\rm T}$), in the marginal one ($c_r=c_{\rm T}$) and in the
post-critical state
 ($c_r<c_{\rm T},\ \vert c_r - c_{\rm T}\vert<<c_{\rm T}$).
In the latter
state, the  mean field $\psi_{1,{\rm int}}$
is no longer destroyed by the peculiar motions of
the stars and, presumably,
affects the
disc properties. One
can use the  linear  response
 function
when  $c_r>c_{\rm T}$
as
in Section 2.
However,
the  linear  approximation
(linear  perturbation analysis) must be abandoned
when we  consider
in this section
the general  equilibrium
state for the marginal
 and
post-critical regimes. We consider perturbations of a small, but finite amplitude.

Let us return to Eq.~(\ref{eq2}).
As follows from the linear theory of density waves and from the above analysis, the most
unstable (critical), as well as the most susceptible wave number is given by
$k^*_{m,\rm uns }=2k_{\rm T}/Q^2$ in
the
marginal state with
$\omega-m\Omega=0$ and $Q=1$, where $k_T=\kappa^2/(2\pi G\sigma),$
$Q=c_r/c_{\rm T},$ and $c_{\rm T}=(\pi G\sigma)/\kappa$ (see Fig.~1). For this
mode, Eq.~(\ref{eq2}) takes the form
\begin{equation}\label{eq3}
    \frac { \sigma_m } { \sigma } \kappa^2
    = ( - c^2_r k_ { m, \rm uns } ^ { * 2 } +
    2\pi G \sigma |k_ { m, \rm uns } ^ *|) \frac { \sigma_m } {
     \sigma } + \psi_ { 1, \rm ext } k_{m,\rm uns}^{*2}\;.
\end{equation}
Using $k_ { m, \rm uns } =\kappa / c_{\rm T}$, canceling $\kappa^2$
 and after some algebra, Eq.~(\ref{eq3}) can be reduced to
\begin{equation}\label{eq4}
    \frac { \sigma_m } { \sigma } (c^2_r - c_{\rm T}^2) = \psi_ {
    1, \rm ext }\;.
\end{equation}

 Equation (\ref{eq4}) correctly describes the susceptibility of
the
 critical mode for $c_r>c_{\rm T}$ but does not yield spontaneous
 phase transition (at $\psi_ { 1, \rm ext } =0$). Furthermore, it does not
 describe the susceptibility in the post-critical state $c_r<c_{\rm T}$
 since it has been obtained without taking into account changes in
 the disc's state.

Now we leave the mode analysis for the general equilibrium relations.
We recall  that in fact a stellar disc is a collisionless dynamic system.

The relation between `pressure' perturbation
$\Delta p$ and gravitational potential function $\psi_1$ applicable to static
perturbations over half the Toomre wavelength
(i.e., the critical  scale) has the form
\begin{equation}
{\Delta p \over \sigma } -\psi_ { 1 } =0\;. \label{star}
\end{equation}
Such perturbations can be called frozen.  They satisfy the general
 hydrostatic equilibrium
equation
for a system in an external field
 known from statistical physics (Landau \& Lifshitz 1976, \S25):

$$\mu_0(P,T)-\Psi(\bf{r}) = {const}\;,$$

where $\mu_0$ and $\Psi$ are the
specific chemical potential (without the field) and field
potential, respectively, $P$ is the `pressure', and $T$
 is the temperature of the system.  $\Psi$ is reduced by the
 centrifugal potential (cf.\ Eq.~(\ref{eqa4}) of the Appendix A that describes
hydrostatic equilibrium of the disc in the pre-critical state).

The `pressure' perturbation near~ the~ critical~ point (i.e.,~ the~ top of
Toomre's  diagram) can be written in a form similar to that used by
Landau in his theory of van der Waals' critical point (Landau \&
Lifshitz 1976, \S152):
\begin{equation}\label{new}
{\Delta p \over \sigma }=c^2_r
{ \sigma_m \over \sigma } + 4 c^2_{\rm T} { \sigma^3_m  \over  \sigma^3}\;.
\end{equation}
The applicability of this expression to stellar
 discs is our assumption,  justified  a  posteriori  by  the  results
 obtained. This  expression follows from general arguments
presented,  e.g.,  by Landau \& Lifshitz  (1976,  \S\S  144,
 152). For static perturbations, this expression results
 in one real and two imaginary solutions for
$\sigma_m$ when
 $c_r>c_{\rm T}$  and three  real  solutions
when $c_r<c_{\rm T}$. Thus, the expression (\ref{new}) leads to a
reasonable physical  behaviour similar to that of a gas--liquid
system, in the sense that the stellar disc maintains hydrostatic
equilibrium near its  marginal state. The expression~(\ref{new})
simply implies the expansion of $\Delta p$ in power of $\sigma_m$,
supposing the inflection (bending) point for equation of state
function of a stellar disk $ p(\sigma)$ (in a marginal  state), as
$\Delta p = (dp/d\sigma)\sigma_m +
(1/n!)(d^3p/d\sigma^3)\sigma_m^3$. That is, the disc's state is
assumed to change  as the disc passes through the critical point,
as we pointed out in Abstract, the inflection (bending) point is
appeared .

The term nonlinear in $\sigma_m$ in the expression~(\ref{new}) can
be understood in the followimg semi-quantitative way. A density
perturbation $\sigma_m$ leads to an increase in stellar velocities
on their orbits by $\Delta v$ due to an increase in the
gravitational force. For static perturbations, the velocity
incerement can be found from centrifugal  balance as
 $\Omega \Delta v=\pi G \sigma_m.$
Thus,
  $\Delta v=\pi G \sigma_m/\Omega \approx 2 \pi G \sigma_m/\kappa $.
The resulting `pressure'  increment $\Delta p$  is given by
 ${(\Delta v)}^2\sigma_m
\approx
4(\pi G \sigma/\kappa)^2 (\sigma_m/\sigma)^2 \sigma_m$,
which is just the nonlinear term of the expression~(\ref{new}).
This term represents an additional `elasticity' (apart from that of
the epicyclic motion,
represented by $c_r$) that arises from the stability of the circular
orbits, associated with
the centrifugal force, because the orbits are assumed to sustain
 gravitational perturbations
for $c_r \leq c_{\rm T}$.

We note that
\begin{eqnarray*}
\psi_1&=&\psi_ { 1, \rm int } + \psi_ { 1, \rm ext }\;,\\
\psi_{1,\rm int}&=&\pi G \sigma_m r_ {\rm corr } = \pi G \sigma_m
{c_{\rm T}\over\kappa } =c_{\rm T}^2 { \sigma_m \over \sigma }\;,
\end{eqnarray*}
which results,
using Eq.~(\ref{star}), in the following equation for
the order parameter $\sigma_m$:
\begin{equation}\label{eq5}
    \frac { \sigma_m } { \sigma }
\left[ c^2_r - c_{\rm T}^2 + 4 c_{\rm T}^2\left(\frac{\sigma_m}{\sigma}\right)^2
\right] = \psi_ { 1, \rm ext }\; .
\end{equation}

Equation~(\ref{eq5})
has a standard form for an equation describing
phase equilibrium in Landau's theory of the phase transition of the
second kind (Landau \& Lifshitz 1976). This equation yields the ratio
controlling the  susceptibility $\partial \sigma_m / \partial \psi_
{ 1, \rm ext } $ in the pre-critical state ($c_r  > c_{\rm T}$),
\begin{equation}\label{eq6}
\frac{\partial\sigma_m}{\partial\psi_ {1,\rm ext}}=\chi
=\frac{\sigma}{c^2_r - c^2_{\rm T}}\;,
\end{equation}
together with expressions for the dependence of the order parameter on
$c^2_r-c^2_{\rm T}$ (if evaluated
at $\psi_{ 1,\rm ext }=0$, see Eq.~(\ref{eq7})
 below) and on the external field (if evaluated at $c_r=c_{\rm T}$; note that the
result remains non-singular),
\[
  \sigma_m\sim\psi_{1,\rm ext}^{1/\delta}\;.
\]

Thus, there are three critical exponents, $\gamma=1,\ \beta=1/2$ and
$\delta=3.$ These are not independent but related by Widom's ratio
(Landau \& Lifshitz 1976, \S148),  $\beta\delta=\beta+\gamma$.

For $c_r<c_{\rm T}$, the susceptibility has to be found for a non-zero value of
$\sigma_m$. For this purpose the bracket in Eq.~(\ref{eq5}) should be put equal to zero.
For $c_r< c_{\rm T}$ we obtain a real solution of the type $\sigma_m/\sigma\sim{(c_{\rm
T}^2-c_r^2)}^{\beta}$
which is given by
\begin{equation}\label{eq7}
   \left( { \sigma_m\over { \sigma }} \right) ^2=
   - {{ c_r^2-c_{\rm T}^2 } \over { 4c_{\rm T}^2 }}\;.
\end{equation}

For $c_r>c_{\rm T}$,  Eq.~(\ref{eq7})
 formally  gives  purely
 imaginary  density perturbations,
implying
that static
 density perturbations of  finite  amplitude  cannot  exist  in  this
 case.

Differentiating Eq.~(\ref{eq5}) and using Eq.~(\ref{eq7}),
we find the susceptibility in the
post-critical case as
\begin{eqnarray}\label{eq8}
\frac{\partial\sigma_m}{\partial\psi_{ 1,\rm ext}}=\chi
&=&\frac{\sigma}{c_r^2-c_{\rm T}^2+12c_{\rm T}^2(\sigma_m/\sigma)^2}\nonumber\\
&=&-{\sigma\over 2 ( c_r^2-c_{\rm T}^2 ) }\;.
\end{eqnarray}
Comparing Eqs (\ref{eq6}) and (\ref{eq8}), we see that the
susceptibility suffers a jump at $c_r=c_{\rm T}$ as it should at a
phase transition.

Finally, the exponents follow as $\beta=1/2,\  \gamma=1,\ \delta=3,$
and $\nu=1 / 2.$ The scalings of Eqs~(\ref{eq6})--(\ref{eq8}) agree
with the results of Landau--Weiss's standard theory (Landau \&
Lifshitz 1976).

\section{Conclusions}
In the spirit of Landau's theory of the van der Waals critical point
 and the generic nature
of critical phenomena, we have shown the possibility of
critical behaviour of galactic stellar discs. The disc undergoes a
phase transition when the stellar velocity dispersion $c_r$ is equal
to Toomre's critical value, $c_{\rm T}$. It appears that the
transition leads to the formation of a fundamental spiral structure.
We have shown that well-known features of such phase transitions
result in an essentially new mechanism for the spiral structure
formation. Similarly to a thermodynamic system undergoing
the  gas--liquid phase transition at a
critical  point,  a  marginally
 stable  collisionless  stellar disc can,
while retaining (or tending to retain)
its hydrostatic equilibrium, exhibit a critical  behaviour  that
 leads to its primary spatial inhomogeneity which can further develop
 into a spiral structure. This is the main
point of our approach which takes advantage of the theory of critical
phenomena; linear results are a part of our analysis, but they are
not sufficient for it.

Numerical experiments of Hohl (1975, 1975a) demonstrate that a
stellar disc embedded in a bulky  halo appears to be rather
changeable at $Q=Q_{\rm cr}$. Processes different from the bar
instability `thermalize' the disc quickly (practically
instantaneously), heating it up to $Q\approx(2$--$3)Q_{\rm cr}$.
At the initial stage of the process, the disc is populated by a
set of stellar splinters and  crescents, and a few large-scale
spiral branches develop later. These structures eventually reduce
to decaying worm-type features because of the `heating' most
likely resulting from the instability of individual stellar orbits
due to resonances. This demonstrates that both dynamical
instabilities and fluctuations (critical and, for example, induced
by embedding bulky masses) can be superimposed on the process
wherein an order emerges. This scenario is also confirmed by
numerical calculations  of critical dynamics in a model
one-dimensional self-gravitating system in a marginal state
(Ivanov 1992, 1993).

 `Hot'
discs cannot have quasistatic  structures, and this fact corroborates
our theory of phase transitions.
Our theory is only applicable to early stages of the simulations
discussed above before instabilities have developed.
There  are many  effects  that can
 halt the ordering process, so it can occur only in some galaxies.

We have considered only compressional perturbations. But the last
 term in the expression~(\ref{new}) for $\Delta p/\sigma$
can be compared with
a change in the stellar velocity driven by gravitational
perturbations during the phase transition, namely with the drift
correction to the chaotic stellar velocity, whose azimuthal component
 is given by $|\kappa^{-1}\partial\psi_{1,\rm int}/\partial r|$ and
the radial component by $|(\kappa r)^{-1} \partial\psi_{1,\rm
int}/\partial\varphi|$ (Ivannikova \& Maksumov 1994).  The square of
the drift correction to the velocity is given by
$|\kappa^{-1}\partial\psi_{1,\rm int}/\partial r|^2 + |(\kappa
r)^{-1}\partial\psi_{1,\rm int }/\partial\varphi|^2$. Using
$\psi_{1,\rm int}=2\pi G\sigma_m/|k^*|$, $k^*_{m,\rm
uns}=\kappa/c_{\rm T}$, $|\kappa^{-1}\partial\psi_{1,\rm
int}/\partial r| =2c_{\rm T}(\sigma_m/\sigma)\cos\delta$ and $(\kappa
r)^{-1}\partial\psi_{1,\rm int}/\partial\varphi|= 2c_{\rm T}
(\sigma_m/\sigma)\sin\delta$, the drift correction squared can be
expressed as $4c_{\rm T}^2(\sigma_m/\sigma)^2.$
This estimates confirms the possibility of translational and shearing
effects in addition to the compressional one.

\section*{Acknowledgments}
We are  grateful  to  Anvar Shukurov for his assistance.


 \section{Appendix A: The derivation of Eq.~(1) for density perturbations in a
      stellar disc. The drift perturbations}\label{app}

Linearized hydrodynamic equations for the density perturbation
$\sigma_m$ and the radial and azimuthal bulk velocities $u$ and
$v$ are given by (Rohlfs 1977, \S5.5)
\begin{equation}\label{eqa1}
      \frac { \partial \sigma_m } { \partial t } + \Omega\frac { \partial
 \sigma_m } { \partial \varphi } + \frac { 1 } { r }\left[  \frac {
\partial } {\partial r } ( r \sigma u) + \frac { \partial } {
 \partial \varphi } ( \sigma v )\right] = 0\;,
\end{equation}
\begin{equation}\label{eqa2}
 \frac { \partial u } { \partial t } + \Omega\frac { \partial u } {
 \partial \varphi } - 2\Omega v = - \frac { c^2_r } { \sigma }\,
 \frac { \partial \sigma_m } { \partial r } + \frac { \partial \psi_1
 } { \partial r }\;,
 \end{equation}
\begin{equation}\label{eqa3}
\frac { \partial v } { \partial t } + \Omega\frac { \partial v } {
 \partial \varphi } + \frac { \kappa^2 } { 2\Omega } u = - \frac { 1 } { r }
 \left(
\frac { c^2_\varphi } { \sigma }\, \frac { \partial \sigma_m } {
 \partial \varphi } - \frac { \partial \psi_1 } { \partial \varphi }
 \right) .
\end{equation}
Assuming that the perturbed quantities (including
 $\psi_1$) are proportional to $\exp[i(\omega t - m
 \varphi + k r)]$ and reserving for amplitudes the
notation used for the respective perturbed quantities, we transform
these equations first similarly to Lin \& Shu (1964) who considered
tightly wound perturbations.
Equations (\ref{eqa1}), (\ref{eqa2}) and (\ref{eqa3}) reduce to
\begin{equation}
      i ( \omega - m \Omega ) \sigma_m + i \sigma k u+{d \sigma\over dr}
       u -i \sigma k_ \varphi v=0\;,\label{eqa1p}
\end{equation}
\begin{equation}
 i\frac { kc^2_r } { \sigma } \sigma_m + i ( \omega - m\Omega ) u - 2
 \Omega v=i k \psi_1\;,\label{eqa2p}
\end{equation}
\begin{equation}
 i ( \omega - m\Omega ) v + \frac { \kappa^2 } { 2\Omega } u =
      i\frac { k_\varphi c^2_\varphi } { \sigma } \sigma_m
-ik_\varphi \psi_1\;.  \label{eqa3p}
\end{equation}
Equation (\ref{eqa3p}) yields for $v$:
\[
  v=i\frac{\kappa^2}{2\Omega(\omega-m\Omega)} u +
      \frac{k_\varphi c^2_\varphi\sigma_m/\sigma-k_\varphi\psi_1}{\omega-m\Omega }\;.
\]
Using this expression in Eq.~(\ref{eqa2p}), we  find $u$ and $v$ as
\begin{eqnarray*}
u&=&{{\omega-m\Omega}\over {{(\omega-m\Omega)}^2-\kappa^2}}\\
&&\mbox{}\!\!\times\!\left[-k \left(c^2_r { \sigma_m\over \sigma}- \psi_1\right)-
        \frac{2i\Omega}{\omega-m\Omega}
        k_\varphi \left(c^2_\varphi\frac{\sigma_m}{\sigma}-\psi_1\right)\right];\\
v&=&{{i \kappa^2/2\Omega}\over {{(\omega-m\Omega)}^2-\kappa^2}}\\
&&\mbox{}\!\!\times\!\left[-k\left(c^2_r{\sigma_m\over\sigma}-\psi_1\right)
        - i{2\Omega\over{\omega-m\Omega}}
        k_\varphi\left(c^2_\varphi {\sigma_m\over \sigma}-
        \psi_1\right)\right]\\
&&\mbox{}+{k_\varphi \over{
          \omega-m\Omega} } \left(c^2_\varphi {\sigma_m\over \sigma}-
        \psi_1\right).
\end{eqnarray*}
Using $c^2_\varphi=( \kappa^2/2\Omega)c^2_r$, we find from Eq.~(\ref{eqa1p}):
\begin{eqnarray*}
\frac{\sigma_m}{\sigma}&=&\frac{1}{[(\omega-m\Omega)^2-\kappa^2](\omega-m\Omega)^2}\\
&&\mbox{}\times\left[
        (k^2+k_\varphi^2)(\omega-m\Omega)^2
        \left(c^2_r \frac{\sigma_m}{\sigma}-\psi_1\right)
                                                                \right.\\
&&\mbox{}+k_\varphi^2 (\omega-m\Omega)^2 c_r^2\,\frac{1}{2}\,
        \frac{d\ln\Omega}{d\ln r} \,\frac{\sigma_m}{\sigma}\\
&&\mbox{}+2\Omega(\omega-m\Omega)\frac{d\ln\sigma}{dr} k_\varphi
\left(\frac{\kappa^2}{2\Omega}\,c^2_r\,\frac{\sigma_m}{\sigma}-\psi_1\right)\\
&&\mbox{}+ikk_\varphi(\omega-m\Omega)\,r\frac{d\Omega}{dr}\,\psi_1\\
&&\mbox{}\left.-i(\omega-m\Omega)^2\,\frac{d\ln\sigma}{dr}\,k
                \left(c^2_r\frac{\sigma_m}{\sigma}-\psi_1\right)\right] .
\end{eqnarray*}
Neglecting terms with spatial derivatives of $\sigma$ and $\Omega$,
we obtain Eq.~(\ref{eq1}).

Now consider slow, drift modes. They are basically different from
 the tightly wound ones and can only be nonaxisymmetric. Their
susceptibility is easier to consider in the limit
$\partial/\partial t + \Omega\partial/\partial \varphi =0$.  The
 system (\ref{eqa1})--(\ref{eqa3}) then takes the form
\begin{equation}\label{eqa1pp}
   \frac { \partial } { \partial r } ( r \sigma u) + \frac { \partial } {
      \partial \varphi } ( \sigma v )  = 0\;,
\end{equation}
\begin{equation}\label{eqa2pp}
 2\Omega v = \frac { c^2_r } { \sigma }\, \frac { \partial \sigma_m }
 { \partial r } - \frac { \partial \psi_1 } { \partial
 r}\;,
\end{equation}
\begin{equation}\label{eqa3pp}
 \frac { \kappa^2 } { 2\Omega }\, u = - \frac { 1 } { r }\left(
 \frac { c^2_r } { \sigma } \,\frac { \partial \sigma_m } { \partial
 \varphi } - \frac { \partial \psi_1 } { \partial \varphi } \right).
\end{equation}
In the limit  $u,v\to0$ we obtain an expression used to determine the
susceptibility of the drift modes:
\begin{equation}\label{eqa4}
 c^2_r\,\frac { \sigma_m } { \sigma } - \psi_1=0\;.
\end{equation}
Using $\psi_1=\psi_{1,\rm int}+\psi_{1,\rm ext}$ and
$\psi_{1,\rm int}=2\pi G\sigma_m/|k^*|$ with
$k^*=4k_{\rm T}$, $k^*={(k^2+k_\varphi^2)}^{1/2}$, we obtain
Eq.~(\ref{eq6})  for the susceptibility $\chi$.


\end{document}